\documentclass[traditabstract]{aa}

\usepackage{graphicx}
\usepackage{txfonts}

\newcommand{\teff}{\mbox{$T_{\rm eff}$}}
\newcommand{\logg}{\mbox{$\log g$}}
\newcommand{\vsini}{\mbox{$v \sin I$}}
\newcommand{\mictrb}{\mbox{$\xi_{\rm t}$}}
\newcommand{\mactrb}{\mbox{$v_{\rm mac}$}}

\newcommand{\etal}{et\,al.}

\begin{document}

\title{Three irradiated and bloated hot Jupiters:}
\subtitle{WASP-76b, WASP-82b \&\ WASP-90b}

\author{
R.G. West\inst{1}  \and
J.-M. Almenara\inst{2} \and
D.R. Anderson\inst{3} \and
F. Bouchy\inst{4} \and
D. J. A. Brown\inst{5} \and 
A. Collier Cameron\inst{5}  \and
M. Deleuil\inst{2} \and
L. Delrez\inst{6}  \and
A. P. Doyle\inst{3} \and
F. Faedi\inst{1}  \and
A. Fumel\inst{6}  \and
M. Gillon\inst{6}  \and
G. H\'ebrard\inst{4} \and
C. Hellier\inst{3} \and
E. Jehin\inst{6} \and
M. Lendl\inst{7}  \and
P.F.L. Maxted\inst{3}  \and
F. Pepe\inst{7} \and
D. Pollacco\inst{1}  \and
D. Queloz\inst{7,8}  \and
D. S\'egransan\inst{7}  \and
B. Smalley\inst{3}  \and
A.M.S. Smith\inst{3,9}  \and
A.H.M.J. Triaud,\inst{7,10}\thanks{Fellow of the Swiss National Science Foundation}  \and
S. Udry\inst{7}}

\institute{Department of Physics, University of Warwick, Coventry CV4 7AL, UK 
\and Aix Marseille Universit\'e, CNRS, LAM (Laboratoire d'Astrophysique de Marseille) UMR 7326, 13388, Marseille, France 
\and Astrophysics Group, Keele University, Staffordshire, ST5 5BG, UK
\and Institut d'Astrophysique de Paris, UMR 7095 CNRS, Universit\'e Pierre \&\ Marie Curie, France; Observatoire de Haute-Provence, CNRS/OAMP, 04870, St Michel l'Observatoire, France
\and SUPA, School of Physics and Astronomy, University of St.\ Andrews, North Haugh,  Fife, KY16 9SS, UK
\and Institut d'Astrophysique et de G\'eophysique, Universit\'e de Li\`ege, All\'ee du 6 Ao\^ut, 17, Bat. B5C, Li\`ege 1, Belgium 
\and Observatoire astronomique de l'Universit\'e de Gen\`eve
51 ch. des Maillettes, 1290 Sauverny, Switzerland
\and Cavendish Laboratory, J J Thomson Avenue, Cambridge, CB3 0HE, UK
\and N.~Copernicus Astronomical Centre, Polish Academy of Sciences, Bartycka 18, 00-716 Warsaw, Poland
\and Department of Physics and Kavli Institute for Astrophysics \&\ Space Research, Massachusetts Institute of Technology, Cambridge, MA 02139, USA}

\titlerunning{WASP-76b, WASP-82b \&\ WASP-90b}
\authorrunning{West et al.}

\abstract{We report three new transiting hot-Jupiter planets discovered from the
WASP surveys combined with radial velocities from OHP/SOPHIE and Euler/CORALIE and photometry from Euler and TRAPPIST.  All three planets
are inflated, with radii 1.7--1.8 R$_{\rm Jup}$.  All orbit hot stars,
F5--F7, and all three stars have evolved, post-MS radii (1.7--2.2
R$_{\sun}$).  Thus the three planets, with orbits of 1.8--3.9 d, are
among the most irradiated planets known.  This reinforces the
correlation between inflated planets and stellar irradiation.}


\maketitle

\section{Introduction}
The naive expectation that a Jupiter-mass planet would have a
one-Jupiter radius has been replaced by the realisation that many of
the hot Jupiters found by transit surveys have inflated radii.
Planets as large as $\sim$\,2 R$_{\rm Jup}$ have been found (e.g.\ WASP-17b,
Anderson \etal\ 2010; HAT-P-32b, Hartman \etal\ 2011).

It is also apparent that inflated planets are found preferentially
around hot stars.  For example Hartman \etal\ 2012 reported three new
HAT-discovered planets, with radii of 1.6--1.7 R$_{\rm Jup}$, all
transiting F stars.  Here we continue this theme by announcing three
new hot Jupiters, again all inflated and all orbiting F stars.

For a discussion of the radii of transiting exoplanets we refer the
reader to the recent paper by Weiss \etal\ (2013).  It seems clear
that stellar irradiation plays a large role in inflating hot Jupiters,
since no inflated planets are known that receive less than 2 $\times$
10$^{8}$ erg s$^{-1}$ cm$^{-2}$ (Miller \&\ Fortney 2011; Demory
\&\ Seager 2011). There is also an extensive literature discussing
other mechanisms for inflating hot Jupiters, such as tidal dissipation
(e.g.\ Leconte \etal\ 2010, and references therein) and Ohmic
dissipation (e.g.\ Batygin \&\ Stevenson 2010).

\section{Observations}
The three transiting-planet systems reported here are near the
equator, and so have been observed by both the SuperWASP-North camera
array on La Palma and by WASP-South at Sutherland in South Africa.
Our methods all follow closely to those in previous WASP discovery
papers.  The WASP camera arrays are described in Pollacco
\etal\ (2006) while our planet-hunting methods are described in
Collier-Cameron \etal\ (2007a) and Pollacco \etal\ (2007).

Equatorial WASP candidates are followed up by obtaining radial
velocities using the SOPHIE spectrograph on the 1.93-m telescope at
OHP (as described in, e.g., H\'ebrard \etal\ 2013) and the CORALIE
spectrograph on the 1.2-m Euler telescope at La Silla (e.g., Triaud
\etal\ 2013). Higher-quality lightcurves of transits are obtained using 
EulerCAM on the 1.2-m telescope (e.g., Lendl \etal\ 2013) and the
robotic TRAPPIST photometer at La Silla (e.g., Gillon \etal\ 2013).  
The observations for our three new planets are listed in Table~1. 

\begin{table}
\caption{Observations\protect\rule[-1.5mm]{0mm}{2mm}}  
\begin{tabular}{lcr}
\hline 
Facility & Date &  \\ [0.5mm] \hline
\multicolumn{3}{l}{{\bf WASP-76:}}\\  
SuperWASP-North & 2008 Sep--2010 Dec & 12\,800 points \\ 
WASP-South & 2008 Jul--2009 Dec & 7700 points \\
OHP/SOPHIE & 2011 Sep--2011 Dec &  9  RVs \\
Euler/CORALIE  & 2012 Feb--2012 Dec  &   8 RVs \\
TRAPPIST & 2011 Nov 06 &  $I$ filter \\
TRAPPIST & 2012 Aug 25 &  $I$ filter \\
EulerCAM  & 2012 Oct 13 & Gunn $r$ filter \\ 
TRAPPIST & 2012 Oct 31 &  $I$ filter \\
TRAPPIST & 2012 Nov 20 &  $I$ filter \\ [0.5mm] 
\multicolumn{3}{l}{{\bf WASP-82:}}\\  
SuperWASP-North & 2008 Oct--2011 Feb & 15\,100  points \\ 
WASP-South & 2008 Oct--2010 Jan & 8600 points \\
OHP/SOPHIE & 2011 Dec--2012 Feb &  8  RVs \\ 
Euler/CORALIE  & 2012 Feb--2013 Mar  &   20 RVs \\
EulerCAM  & 2012 Nov 20 & Gunn $r$ filter \\ [0.5mm] 
\multicolumn{3}{l}{\bf WASP-90:}\\  
SuperWASP-North & 2004 May--2010 Oct & 40\,800 points \\ 
WASP-South & 2008 Jun--2009 Oct & 12\,200 points \\
Euler/CORALIE  & 2011 Oct--2012 Sep  &   15 RVs  \\
TRAPPIST & 2012 Jun 03 & $I+z$ filter \\
EulerCAM  & 2012 Jul 28 & Gunn $r$ filter \\ 
TRAPPIST & 2012 Sep 13  &  $I+z$ filter \\ 
EulerCAM  & 2013 Jun 10 & Gunn $r$ filter \\ [0.5mm] 
\end{tabular} 
\end{table}

\section{The host stars}
The stellar parameters for WASP-76, WASP-82 and WASP-90 were derived from the
co-added RV spectra using the methods given in Doyle \etal\ (2013). The
excitation balance of the Fe~{\sc i} lines was used to determine the effective
temperature (\teff). The surface gravity (\logg) was determined from the
ionisation balance of Fe~{\sc i} and Fe~{\sc ii}. The Ca~{\sc i} line at
6439{\AA} and the Na~{\sc i} D lines were also used as \logg\ diagnostics.
Values of microturbulence (\mictrb) were obtained by requiring a null-dependence
on abundance with equivalent width. The elemental abundances were determined
from equivalent width measurements of several unblended lines. The quoted error
estimates include that given by the uncertainties in \teff\ and \logg, as well
as the scatter due to measurement and atomic data uncertainties. The projected
stellar rotation velocity (\vsini) was determined by fitting the profiles of
several unblended Fe~{\sc i} lines. Macroturbulence was obtained from the
calibration by Bruntt et\,al.\ (2010).

For WASP-76, the rotation rate ($P = 17.6 \pm 4.0$~d) implied by the
{\vsini} (assuming that the spin axis is perpendicular to us) gives a gyrochronological age of $5.3^{+6.1}_{-2.9}$~Gyr
using the Barnes (2007) relation.  The lithium age of
several Gyr, estimated using results in Sestito \& Randich (2005), is
consistent. For WASP-90, the rotation rate ($P = 11.1 \pm 1.6$~d)
implied by the {\vsini} gives a gyrochronological age of
$4.4^{+8.4}_{-2.4}$~Gyr.  The {\teff} of this star is close to the
lithium-gap (B{\"o}hm-Vitense, 2004), and thus the lack of any
detectable lithium in this star does not provide a usable age
constraint. WASP-82 is too hot for reliable gyrochronological or
lithium ages.

We list in Tables 2--4 the proper motions of the three stars from the
UCAC4 catalogue (Zacharias \etal\ 2013). These are compatible with the
stars being from the local thin-disc population. We also searched the
WASP photometry of each star for rotational modulations by using a
sine-wave fitting algorithm as described by Maxted \etal\ (2011). For
none of the three stars was a significant periodicity found ($<$\,1
mmag at 95\%-confidence).

\section{System parameters}
The radial-velocity and photometric data (Table~1) were combined in a
simultaneous Markov-chain Monte-Carlo (MCMC) analysis to find the
system parameters (see Collier Cameron \etal\ 2007b for an account of
our methods). For limb-darkening we used the 4-parameter law from
Claret (2000), and list the resulting parameters in Tables 2--4.  

For WASP-76b and WASP-82b the radial-velocity data imply circular
orbits with eccentricities less than 0.05 and 0.06 respectively.
WASP-90 is a fainter star and WASP-90b is a lower-mass planet, so,
while the data are again compatible with a circular orbit, the limit
on the eccentricity is weaker at 0.5.  For all three we enforced a
circular orbit in the MCMC analysis (see Anderson \etal\ 2012 for the
rationale for this).  One of the WASP-82 RVs was taken during transit,
and this point was given zero weight in the analysis.  To translate
transit and radial-velocity information (which give stellar density)
into the star's mass and radius we need one additional mass--radius
constraint.  Here we use the calibration presented by Southworth
(2011).

The fitted parameters were thus $T_{\rm c}$,
$P$, $\Delta F$, $T_{14}$, $b$, $K_{\rm 1}$, where $T_{\rm c}$ is the
epoch of mid-transit, $P$ is the orbital period, $\Delta F$ is the
fractional flux-deficit that would be observed during transit in the
absence of limb-darkening, $T_{14}$ is the total transit duration
(from first to fourth contact), $b$ is the impact parameter of the
planet's path across the stellar disc, and $K_{\rm 1}$ is the stellar
reflex velocity semi-amplitude.   The resulting fits are reported in Tables 2 to 4.

\section{Discussion}
The three host stars, WASP-76, WASP-82 and WASP-90, are all F stars
with temperatures of 6250--6500 K. Their metallicities ([Fe/H] =
0.1--0.2) and space velocities are compatible with the local thin-disk
population.  The stellar densities derived from the MCMC analysis,
along with the temperatures from the spectral analysis, are shown on a
modified H--R diagram in Fig.~4.  All three stars have inflated radii
($R_{\ast}$ = 1.7--2.2 R$_{\sun}$) and thus appear to have evolved
significantly. The indicated ages of $\sim$\,2 Gyr are compatible with
the estimates from gyrochronology (Section~2).

The three planets also all have inflated radii (1.7--1.8 R$_{\rm Jup}$) 
and are thus bloated hot Jupiters.  This is likely related to the high
temperature and large radii of the host stars.  Weiss \etal\ (2013)
have shown that the radii of hot-Jupiter planets correlates well with
their irradiation.  Our three new planets fit their relationship, and
are all at the extreme high-irradiation, high-radius end (their
locations on Fig 14 of Weiss \etal\ being at fluxes 5.1, 5.2 and  2.9
$\times 10^{9}$ erg cm$^{-2}$, s$^{-1}$, and radii 20.1, 18.4 and 19.6
R$_{\rm Earth}$ for WASP-76b, WASP-82b and WASP-90b respectively).  One caution, however, is that we have a bias against finding un-bloated hot Jupiter around evolved, high-radius stars, since the transit depths would be low.

Of particular note is that, at $V$ = 9.5 and $R_{\rm P}$ = 1.8 R$_{\rm
  Jup}$, WASP-76 is now the brightest known star transited by a planet
larger than 1.5 R$_{\rm Jup}$.  WASP-82 is not far behind at $V$ =
10.1 and $R_{\rm P}$ = 1.7 R$_{\rm Jup}$, comparable to WASP-79 ($V$ =
10.1, $R_{\rm P}$ = 1.7 R$_{\rm Jup}$; Smalley \etal\ 2012, and KOI-13
($V$ = 10.0, $R_{\rm P}$ = 1.8 R$_{\rm Jup}$; Santerne \etal\ 2012).
Thus the new discoveries will be useful for studying bloated hot
Jupiters.  For example, Triaud (2011) suggests that the orbital
inclinations of hot Jupiters are a function of system age.  Given that
radius changes of evolved systems give age constraints, WASP-76 and
WASP-82 will be good systems for testing this idea.  WASP-82 has a
relatively small \vsini\ (Table~3) for its spectral type, which could
indicate a mis-aligned orbit.

\begin{figure}
\hspace*{-5mm}\includegraphics[width=10cm]{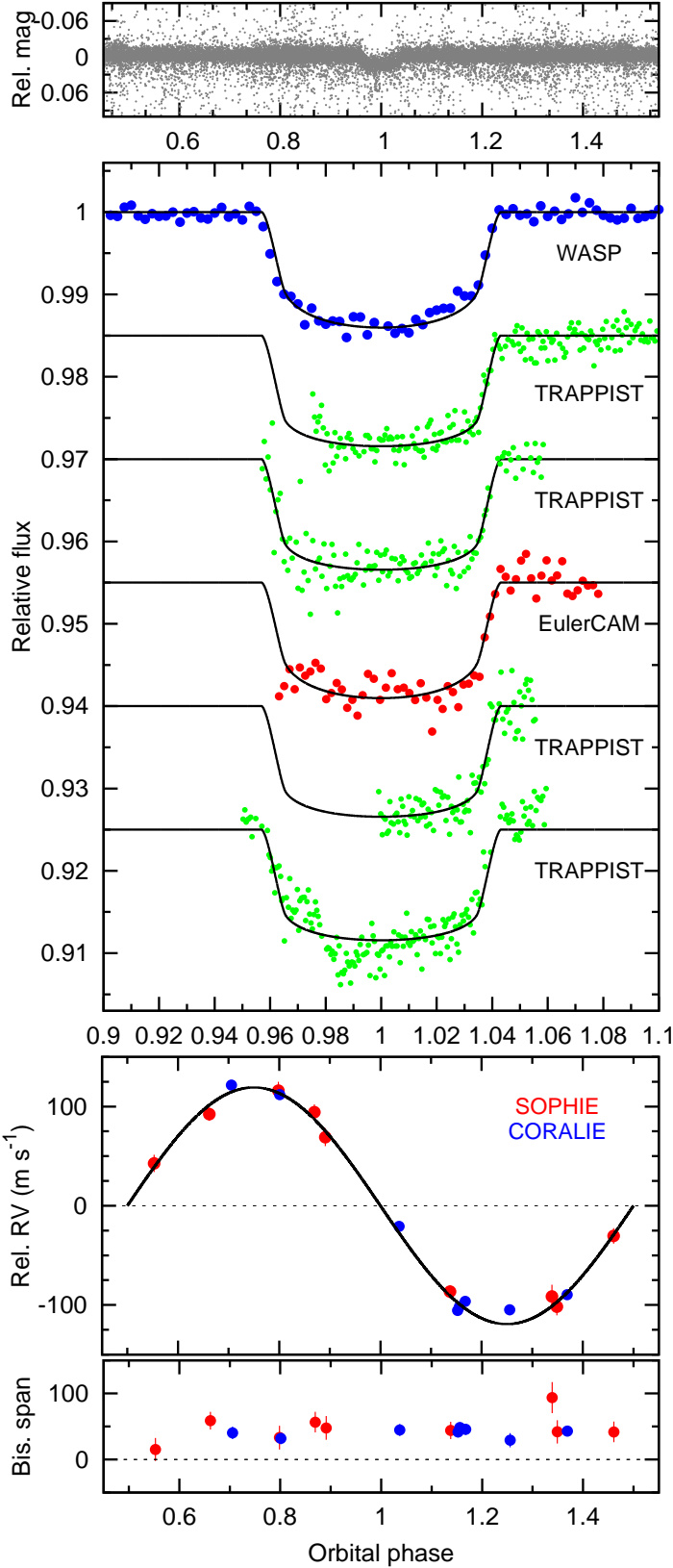}\\ [-2mm]
\caption{WASP-76b discovery data: (Top) The WASP data folded on the 
transit period. (Second panel) The binned WASP data with (offset) the
follow-up transit lightcurves (ordered from the top as 
in Table~1) together with the 
fitted MCMC model.  (Third) The SOPHIE and CORALIE radial
velocities with the fitted model.
(Lowest) The bisector spans; the absence of any correlation with
radial velocity is a check against transit mimics.}
\end{figure}

\begin{table}
\caption{System parameters for WASP-76.}  
\begin{tabular}{lc}
\multicolumn{2}{l}{BD+01 316}\\  
\multicolumn{2}{l}{1SWASP\,J014631.86+024202.0}\\
\multicolumn{2}{l}{2MASS 01463185+0242019}\\
\multicolumn{2}{l}{RA\,=\,01$^{\rm h}$46$^{\rm m}$31.86$^{\rm s}$, 
Dec\,=\,+02$^{\circ}$42$^{'}$02.0$^{''}$ (J2000)}\\
\multicolumn{2}{l}{$V$ mag = 9.5}  \\ 
\multicolumn{2}{l}{Rotational modulation\ \ \ $<$\,1 mmag (95\%)}\\
\multicolumn{2}{l}{pm (RA) 46.6\,$\pm$\,0.7 (Dec), --39.9\,$\pm$\,0.6 mas/yr}\\
\hline
\multicolumn{2}{l}{Stellar parameters from spectroscopic analysis.\rule[-1.5mm]{
0mm}{2mm}} \\ \hline 
Spectral type & F7 \\
$T_{\rm eff}$ (K)      & 6250 $\pm$ 100  \\
$\log g$      & 4.4 $\pm$ 0.1 \\
$\xi_{\rm t}$ (km\,s$^{-1}$)    & 1.4 $\pm$ 0.1 \\
 \mactrb (km\,s$^{-1}$) & 4.0$\pm$0.3 \\
$v\,\sin I$ (km\,s$^{-1}$)     & 3.3 $\pm$ 0.6 \\
{[Fe/H]}   &  +0.23 $\pm$ 0.10 \\
log A(Li)  &  2.28  $\pm$ 0.10 \\
Distance   &   120 $\pm$ 20 pc \\ [0.5mm] \hline
\multicolumn{2}{l}{Parameters from MCMC analysis.\rule[-1.5mm]{0mm}{3mm}} \\
\hline 
$P$ (d) &  1.809886 $\pm$ 0.000001 \\
$T_{\rm c}$ (HJD)\,(UTC) & 245\,6107.85507 $\pm$ 0.00034\\ 
$T_{\rm 14}$ (d) & 0.1539 $\pm$ 0.0008\\ 
$T_{\rm 12}=T_{\rm 34}$ (d) & 0.0154$^{+0.0008}_{-0.0003}$ \\
$\Delta F=R_{\rm P}^{2}$/R$_{*}^{2}$ & 0.01189 $\pm$ 0.00016\\ 
$b$ & 0.14 $^{+0.11}_{-0.09}$ \\
$i$ ($^\circ$) & 88.0 $^{+ 1.3}_{- 1.6}$\\
$K_{\rm 1}$ (km s$^{-1}$) & 0.1193 $\pm$ 0.0018\\ 
$\gamma$ (km s$^{-1}$) & --1.0733 $\pm$ 0.0002\\ 
$e$ & 0 (adopted) ($<$0.05 at 3$\sigma$) \\ 
$M_{\rm *}$ (M$_{\rm \odot}$) & 1.46 $\pm$ 0.07\\ 
$R_{\rm *}$ (R$_{\rm \odot}$) & 1.73 $\pm$ 0.04\\
$\log g_{*}$ (cgs) & 4.128 $\pm$ 0.015\\
$\rho_{\rm *}$ ($\rho_{\rm \odot}$) & 0.286$^{+ 0.008}_{- 0.018}$\\
$T_{\rm eff}$ (K) & 6250 $\pm$ 100\\
$M_{\rm P}$ (M$_{\rm Jup}$) & 0.92 $\pm$ 0.03\\
$R_{\rm P}$ (R$_{\rm Jup}$) & 1.83$^{+0.06}_{-0.04}$\\
$\log g_{\rm P}$ (cgs) & 2.80 $\pm$ 0.02\\
$\rho_{\rm P}$ ($\rho_{\rm J}$) & 0.151 $\pm$ 0.010 \\
$a$ (AU)  & 0.0330 $\pm$ 0.0005\\
$T_{\rm P, A=0}$ (K) & 2160 $\pm$ 40\\ [0.5mm] \hline 

\multicolumn{2}{l}{Errors are 1$\sigma$; Limb-darkening coefficients were:}\\
TRAPPIST $z$:\\
\multicolumn{2}{l}{a1 =    0.683, a2 = --0.349, a3 =  0.565, 
a4 = --0.286}\\ 
EulerCAM $r_{G}$:\\ 
\multicolumn{2}{l}{ a1 =    0.593, a2 = 0.021, a3 =  0.327, 
a4 = --0.215}\\ \hline
\end{tabular} 
\end{table} 

\clearpage

\begin{figure}
\hspace*{-5mm}\includegraphics[width=10cm]{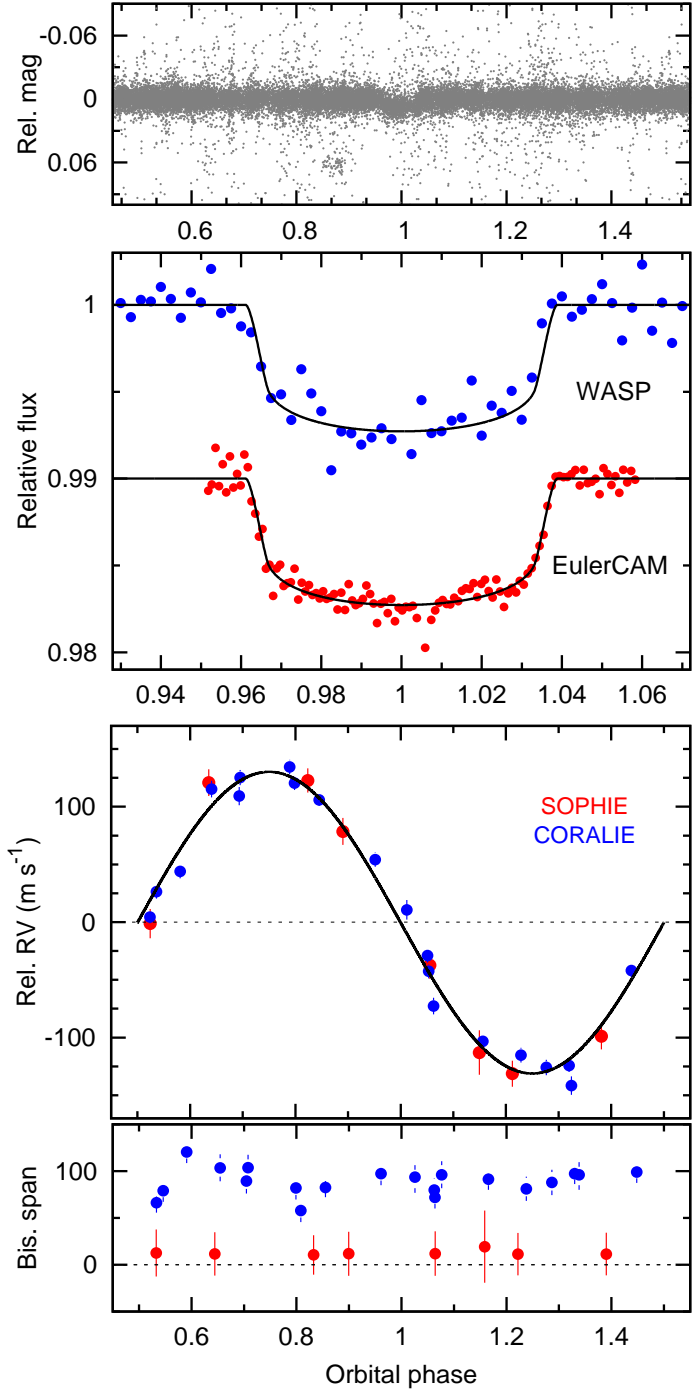}\\ [-2mm]
\caption{WASP-82b discovery data (as for Fig.~1)}
\end{figure}

\begin{table}
\caption{System parameters for WASP-82.}  
\begin{tabular}{lc}
\multicolumn{2}{l}{1SWASP\,J045038.56+015338.1}\\
\multicolumn{2}{l}{2MASS 04503856+0153381}\\
\multicolumn{2}{l}{RA\,=\,04$^{\rm h}$50$^{\rm m}$38.56$^{\rm s}$, 
Dec\,=\,+01$^{\circ}$53$^{'}$38.1$^{''}$ (J2000)}\\
\multicolumn{2}{l}{$V$ mag = 10.1}  \\ 
\multicolumn{2}{l}{Rotational modulation\ \ \ $<$\,0.6 mmag (95\%)}\\
\multicolumn{2}{l}{pm (RA) $-$17.5\,$\pm$\,0.9 (Dec), --17.7\,$\pm$\,0.7 mas/yr}\\
\hline
\multicolumn{2}{l}{Stellar parameters from spectroscopic analysis.\rule[-1.5mm]{
0mm}{2mm}} \\ \hline 
Spectral type & F5 \\
$T_{\rm eff}$ (K)      & 6500 $\pm$ 80  \\
$\log g$      & 4.18 $\pm$ 0.09 \\
$\xi_{\rm t}$ (km\,s$^{-1}$)    & 1.5 $\pm$ 0.1 \\ 
\mactrb  (km\,s$^{-1}$)    &  5.0$\pm$0.3 \\
$v\,\sin I$ (km\,s$^{-1}$)     & 2.6 $\pm$ 0.9 \\
{[Fe/H]}   &  +0.12 $\pm$ 0.11 \\
log A(Li)  &  3.11  $\pm$ 0.08 \\ 
Distance   &   200 $\pm$ 30 pc \\ [0.5mm] \hline
\multicolumn{2}{l}{Parameters from MCMC analysis.\rule[-1.5mm]{0mm}{3mm}} \\
\hline 
$P$ (d) &  2.705782 $\pm$ 0.000003 \\
$T_{\rm c}$ (HJD)\,(UTC) & 245\, 6157.9898 $\pm$ 0.0005\\ 
$T_{\rm 14}$ (d) & 0.2077 $\pm$ 0.0012\\ 
$T_{\rm 12}=T_{\rm 34}$ (d) & 0.0156$^{+0.0012}_{-0.0004}$ \\
$\Delta F=R_{\rm P}^{2}$/R$_{*}^{2}$ &0.00624  $\pm$ 0.00012\\ 
$b$ & 0.16 $^{+0.14}_{-0.11}$ \\
$i$ ($^\circ$) & 87.9 $^{+ 1.4}_{- 1.9}$\\
$K_{\rm 1}$ (km s$^{-1}$) & 0.1307 $\pm$ 0.0019\\ 
$\gamma$ (km s$^{-1}$) & --23.62827 $\pm$ 0.00007\\ 
$e$ & 0 (adopted) ($<$0.06 at 3$\sigma$) \\ 
$M_{\rm *}$ (M$_{\rm \odot}$) & 1.63 $\pm$ 0.08\\ 
$R_{\rm *}$ (R$_{\rm \odot}$) & 2.18 $^{+0.08}_{-0.05}$\\
$\log g_{*}$ (cgs) & 3.973  $^{+0.013}_{-0.02}$ \\
$\rho_{\rm *}$ ($\rho_{\rm \odot}$) & 0.158$^{+ 0.006}_{- 0.014}$\\
$T_{\rm eff}$ (K) & 6490 $\pm$ 100\\
$M_{\rm P}$ (M$_{\rm Jup}$) & 1.24 $\pm$ 0.04\\
$R_{\rm P}$ (R$_{\rm Jup}$) & 1.67$^{+0.07}_{-0.05}$\\
$\log g_{\rm P}$ (cgs) & 3.007 $^{+0.017}_{-0.032}$ \\
$\rho_{\rm P}$ ($\rho_{\rm J}$) & 0.266$^{+0.017}_{-0.029}$ \\
$a$ (AU)  & 0.0447 $\pm$ 0.0007\\
$T_{\rm P, A=0}$ (K) & 2190 $\pm$ 40\\ [0.5mm] \hline 

\multicolumn{2}{l}{Errors are 1$\sigma$; Limb-darkening coefficients were:}\\
EulerCAM $r_{G}$:\\
\multicolumn{2}{l}{ a1 =    0.494, a2 = 0.424, a3 =  --0.266, a4 = 0.0436}\\ \hline
\end{tabular} 
\end{table}

\clearpage

\begin{figure}
\hspace*{-5mm}\includegraphics[width=10cm]{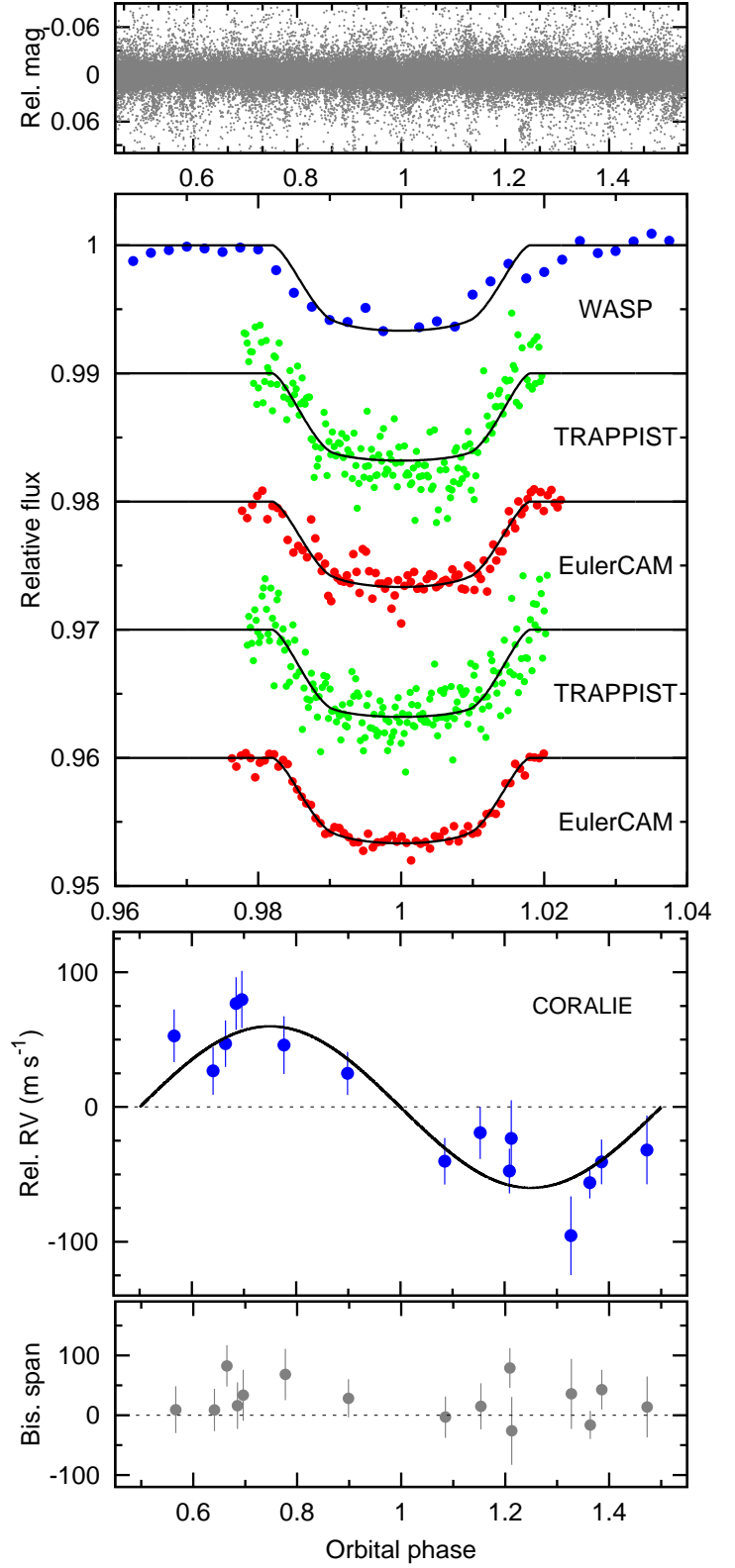}\\ [-2mm]
\caption{WASP-90b discovery data (as for Fig.~1)}
\end{figure}

\begin{table}
\caption{System parameters for WASP-90.}  
\begin{tabular}{lc}
\multicolumn{2}{l}{1SWASP\,J210207.70+070323.7}\\
\multicolumn{2}{l}{2MASS 01463185+0242019}\\
\multicolumn{2}{l}{RA\,=\,21$^{\rm h}$02$^{\rm m}$07.70$^{\rm s}$, 
Dec\,=\,+07$^{\circ}$03$^{'}$23.7$^{''}$ (J2000)}\\
\multicolumn{2}{l}{$V$ mag = 11.7}  \\ 
\multicolumn{2}{l}{Rotational modulation\ \ \ $<$\,1 mmag (95\%)}\\
\multicolumn{2}{l}{pm (RA) $-$10.2\,$\pm$\,1.4 (Dec), 8.1\,$\pm$\,4.3 mas/yr}\\
\hline
\multicolumn{2}{l}{Stellar parameters from spectroscopic analysis.\rule[-1.5mm]{
0mm}{2mm}} \\ \hline 
Spectral type & F6 \\
$T_{\rm eff}$ (K)      & 6440 $\pm$ 130  \\
$\log g$      & 4.32 $\pm$ 0.09 \\
$\xi_{\rm t}$ (km\,s$^{-1}$)    & 1.3 $\pm$ 0.2 \\ 
\mactrb  (km\,s$^{-1}$)    &  4.7$\pm$0.3 \\
$v\,\sin I$ (km\,s$^{-1}$)     & 6.0 $\pm$ 0.5 \\
{[Fe/H]}   &  +0.11 $\pm$ 0.14 \\
log A(Li)  &  $<$1.7  \\
Distance   &   340 $\pm$ 60 pc \\ [0.5mm] \hline
\multicolumn{2}{l}{Parameters from MCMC analysis.\rule[-1.5mm]{0mm}{3mm}} \\
\hline 
$P$ (d) & 3.916243  $\pm$ 0.000003 \\
$T_{\rm c}$ (HJD)\,(UTC) & 245\,6235.5639 $\pm$ 0.0005\\ 
$T_{\rm 14}$ (d) & 0.1398 $\pm$ 0.0022\\ 
$T_{\rm 12}=T_{\rm 34}$ (d) & 0.033 $\pm$ 0.003  \\
$\Delta F=R_{\rm P}^{2}$/R$_{*}^{2}$ & 0.0071 $\pm$ 0.0002\\ 
$b$ & 0.841 $\pm$ 0.013  \\
$i$ ($^\circ$) & 82.1 $\pm$ 0.4 \\
$K_{\rm 1}$ (km s$^{-1}$) & 0.060 $\pm$ 0.006\\ 
$\gamma$ (km s$^{-1}$) & 4.361 $\pm$ 0.0003\\ 
$e$ & 0 (adopted) ($<$0.5 at 3$\sigma$) \\ 
$M_{\rm *}$ (M$_{\rm \odot}$) & 1.55 $\pm$ 0.10\\ 
$R_{\rm *}$ (R$_{\rm \odot}$) & 1.98 $\pm$ 0.09\\
$\log g_{*}$ (cgs) & 4.033 $\pm$ 0.029\\
$\rho_{\rm *}$ ($\rho_{\rm \odot}$) & 0.20 $\pm$ 0.02 \\
$T_{\rm eff}$ (K) & 6430 $\pm$ 130\\
$M_{\rm P}$ (M$_{\rm Jup}$) & 0.63 $\pm$ 0.07\\
$R_{\rm P}$ (R$_{\rm Jup}$) & 1.63 $\pm$ 0.09\\
$\log g_{\rm P}$ (cgs) & 2.73 $\pm$ 0.06\\
$\rho_{\rm P}$ ($\rho_{\rm J}$) & 0.145 $\pm$ 0.027 \\
$a$ (AU)  & 0.0562 $\pm$ 0.0012\\
$T_{\rm P, A=0}$ (K) & 1840 $\pm$ 50\\ [0.5mm] \hline 

\multicolumn{2}{l}{Errors are 1$\sigma$; Limb-darkening coefficients were:}\\
TRAPPIST $I+z$:\\
\multicolumn{2}{l}{a1 =    0.554, a2 = 0.041, a3 =  0.070, 
a4 = --0.086}\\ 
EulerCAM $r_{G}$:\\
\multicolumn{2}{l}{a1 =    0.476, a2 = 0.422, a3 =  --0.226, 
a4 = 0.020}\\ \hline
\end{tabular} 
\end{table}

\clearpage

\begin{figure}
\hspace*{-2mm}\includegraphics[width=9.5cm]{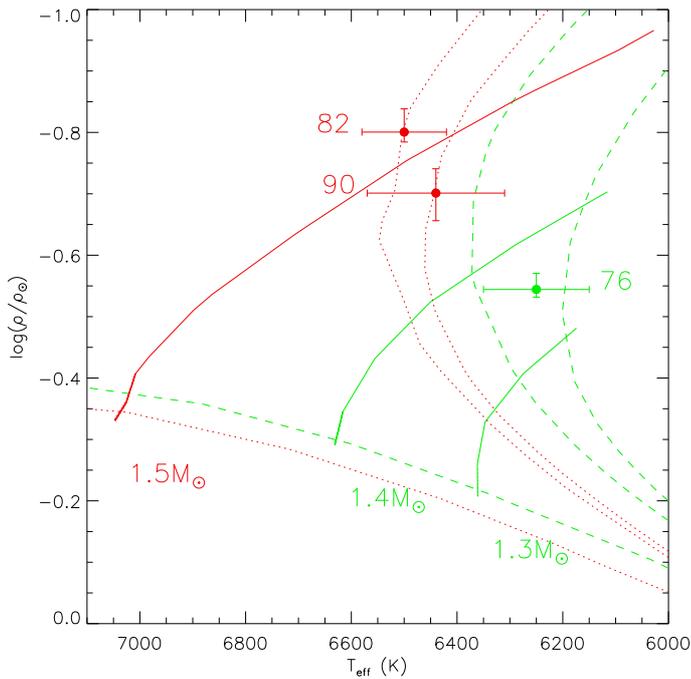}
\caption{Evolutionary tracks on a modified H--R diagram
($\rho$ versus $T_{\rm eff}$).
The green lines are for a metallicity of [Fe/H] = +0.19; 
the dashed lines are isochrones for 0.07, 2.0 and 2.5 Gyr;
the solid lines are mass tracks for 1.3 M$_{\odot}$ and 1.4 M$_{\odot}$. The red lines are for a metallicity of [Fe/H] = +0.1, with the same isochrones, and the mass track for 1.5 M$_{\odot}$.  The models are from Girardi \etal\ (2000).}
\end{figure}

\begin{acknowledgements}
SuperWASP-North is hosted by the Issac Newton Group and the
Instituto de Astrof\'isica de Canarias on La Palma while 
WASP-South is hosted by the South African
Astronomical Observatory; we are grateful for their ongoing
support and assistance. Funding for WASP comes from consortium universities
and from the UK's Science and Technology Facilities Council.
TRAPPIST is funded by the Belgian Fund for Scientific  
Research (Fond National de la Recherche Scientifique, FNRS) under the  
grant FRFC 2.5.594.09.F, with the participation of the Swiss National  
Science Fundation (SNF).  M. Gillon and E. Jehin are FNRS Research  
Associates. A.H.M.J. Triaud is a Swiss National Science Foundation Fellow under grant PBGEP2-145594

\end{acknowledgements}

\end{document}